\makeatletter
\@ifundefined{@parse@version@dash}{%
\def\@parse@version#1{\@parse@version@0#1}
\def\@parse@version@#1/#2/#3#4#5\@nil{%
\@parse@version@dash#1-#2-#3#4\@nil}
\def\@parse@version@dash#1-#2-#3#4#5\@nil{%
  \if\relax#2\relax\else#1\fi#2#3#4 }
}{}
\makeatother

\documentclass[aps,pre,reprint,groupedaddress]{revtex4-2}

\usepackage{graphicx}
\usepackage{amsmath}
\usepackage[utf8]{inputenc}
\usepackage[T1]{fontenc}

\begin{document}


\title{Asymmetric invasion in anisotropic porous media}


\author{Dario Maggiolo}
\email[]{maggiolo@chalmers.se}
\affiliation{Department of Mechanics and Maritime Sciences, Chalmers University of Technology,
Göteborg, SE-412 96, Sweden}
\author{Francesco Picano}
\affiliation{Department of Industrial Engineering, University of Padova, Padova, 35131, Italy}
\author{Federico Toschi}
\affiliation{Department of Applied Physics, Eindhoven University of Technology, Eindhoven 5600 MB, The Netherlands}


\date{\today}

\begin{abstract}
We report and discuss, by means of pore-scale numerical simulations, the possibility of achieving a directional-dependent two-phase flow behaviour during the process of invasion of a viscous fluid into anisotropic porous media with controlled design. 
By customising the pore-scale morphology and heterogeneities with the adoption of anisotropic triangular pillars distributed with quenched disorder,
we observe a substantially different invasion dynamics according to the   direction of fluid injection relative to the medium orientation, that is 
depending if the triangular pillars have their apex oriented (flow-aligned) or opposed (flow-opposing) to the main flow direction.
Three flow regimes can be observed: (i) for low values of the ratio between the macroscopic pressure drop and the characteristic pore-scale capillary threshold, i.e. for $\Delta p_0 /p_c \le 1 $, the fluid invasion dynamics is strongly impeded and the viscous fluid is unable to reach the outlet of the medium, irrespective of the direction of injection; (ii) for intermediate values, $ 1 < \Delta p_0 /p_c \le 2 $,  the viscous fluid reaches the outlet only when the triangular pillars are flow-opposing oriented; (iii) for larger values, i.e for $\Delta p_0/p_c>2$, the outlet is again reached irrespective of the direction of injection.
The porous medium anisotropy induces a lower effective resistance when the pillars are flow-opposing oriented, suppressing front roughening and capillary fingering. 
We thus argue that the invasion process occurs as long as the pressure drop is larger then the macroscopic capillary pressure determined by the front roughness, which in the case of flow-opposing pillars is halved. 
We present a simple approximated model, based on Darcy's assumptions, 
that links the macroscopic effective permeability with the directional-dependent front roughening , to predict the asymmetric invasion dynamics.
This peculiar behaviour opens up the possibility of fabrication of porous capillary valves to control the flow along certain specific directions.
\end{abstract}


\maketitle


\section{Introduction}

Capillary valves are non-mechanical valves that make use of interfacial tension forces to control the fluid flows. They find important applications in microfluidic systems, where such control is crucial for enabling the desired distribution of reagents in order to regulate chemical and biological processes, among which the fabrication of point-of-care diagnostic devices~\cite{arango2020electro,fu2017progress}. 
Examples of capillary valves are, e.g., capillary burst valves, where an abrupt change in geometry provides a capillary resistance that acts as a barrier to the flow along specific directions, and hydrophobic valves, where the wetting properties of the microfluidic walls can be tuned for regulating the fluid motion~\citep{chen2008analysis,olanrewaju2018capillary}.  

Porous elements may be designed and adapted to control the spatial and temporal distribution of the flows for different purposes. As an example, because of their versatile wicking properties, paper-based porous materials are a promising technology for the fabrication of microfluidics~\cite{yang2017based}. Porous media are also used as fuel cell electrodes, where they are hydrophobically treated to better control the water spatial and temporal distribution at the microscale~\cite{nagai2019improving}.
The wetting property of the medium is known to be an important parameter to determine the water flow intensity in soil porous substrates~\cite{doerr2000soil}, 
and it can be tune to obtain materials with directional fluid transport features~\cite{zhao2017directional}.
While flow regulation through the alteration of the porous material wetting properties is under current development, less is know about fluid control by means of the tailoring of the porous microstructure.

The microstructure of a porous medium, intended as the small scale structure characterising its geometrical features, plays an important role in determining the fluid invasion process when a viscous fluid invades a porous medium initially filled with a less viscous one.
The spatial configuration and dynamics of the two-phase interface is determined by the fluid-fluid displacements occurring at the pore scale, the so-called invasion events, which are stochastic, often simultaneous, pore-by-pore invasion processes related to the geometrical disorder of the microstructure.
Among the properties that define a porous microstructure, pore-scale heterogeneities, i.e. small-scale deviations in the regular geometrical patterns, can greatly impact mass transport in porous media, e.g. in diffusion and
surface reactions problems~\cite{haggerty1995multiple}.  
Furthermore, pore morphology, its form and shape, is directly linked to the minimum capillary pressure (capillary threshold) that determines the minimum interfacial energy required for a fluid to invade a pore passing through a 
preceding constrictions (pore throat), because it experiences an increase of surface area. The capillary thresholds to access a pore can be tweaked via pore morphology design in order to e.g. control the emergence of three-dimensional structure in micropillar scaffolds~\cite{yasuga2021fluid}. 
 Anisotropy is know to greatly affect the permeability and dispersion of the medium in single phase flow~\cite{maggiolo2016flow} and it has been observed to have a significant impact on multiphase flow properties of stratified rock formations~\cite{bakhshian2020co2}.

Throughout this paper, for the sake of simplicity, we refer to the injected fluid phase 1 (invading fluid), as to the viscous phase, which presents a higher dynamic viscosity compared to the fluid 2 that initially fills the medium (the displaced fluid), i.e. $\mu_1\gg\mu_2$. 
 In such a situation, the collective dynamics of invasion events is usually referred as stable displacement, since the viscous forces dominates the invading fluid dynamics inducing a lower pressure at the tip of the invading front; consequently, the rougher the interface and the farther downstream the tip is found compared to the average front position, the lower is the tip probability of overcoming the capillary thresholds at the contiguous pore throats, for successive invasion events~\cite{aker2000dynamics,cottin2010drainage}. A stable displacement process thus tends to compact the front and limit the front roughness. For example, in Fig.~\ref{fig1}(b) a stable displacement would promote the configuration depicted in the lower panel rather than the one in represented in the upper one. 

However, such a stabilising mechanism, which mainly depends on the fluid-fluid dynamic viscosity ratio, can be opposed by other relevant factors, such as the instability generated at the pore scale with high values of the advancing contact angle of the invading fluid~\cite{jung2016wettability}, inertial forces~\cite{moebius2014inertial,zhang2020capillary}, or geometrical configurations that provides a positive permeability gradient along the fluid invasion direction~\cite{cottin2010drainage}. The latter mechanism of instability, which promotes the kinetic roughening of the invading front, has been observed both in porous materials reconstructed via X-ray computed tomography~\cite{farzaneh2020pore} and in polydimethylsiloxane (PDMS) microfluidic devices manufactured with controlled pore size gradients~\cite{lu2019controlling}. 
 
In this study we present pore-resolved Lattice Boltzmann simulations to further understand the relationship between pore morphology, pore-scale heterogeneity and two-phase flows. We make use of a simple two-dimensional geometrical configuration, where the porous microstructure is tweaked by introducing anisotropic solid elements (triangular pillars) and defects (missing pillars), to define medium anisotropy and heterogeneity. We show that, by combining this two geometrical elements, it is possible to construct porous capillary valves (i.e. a device where, for a given pressure gradient, the mean flows along one and the other directions are substantially different), whose versatile functioning can be adapted in order to regulate the flow retention along a specific direction.

\section{A porous capillary valve with asymmetric fluid conductance~\label{sec:valve}}

\begin{figure}
\includegraphics[width=0.99\linewidth]{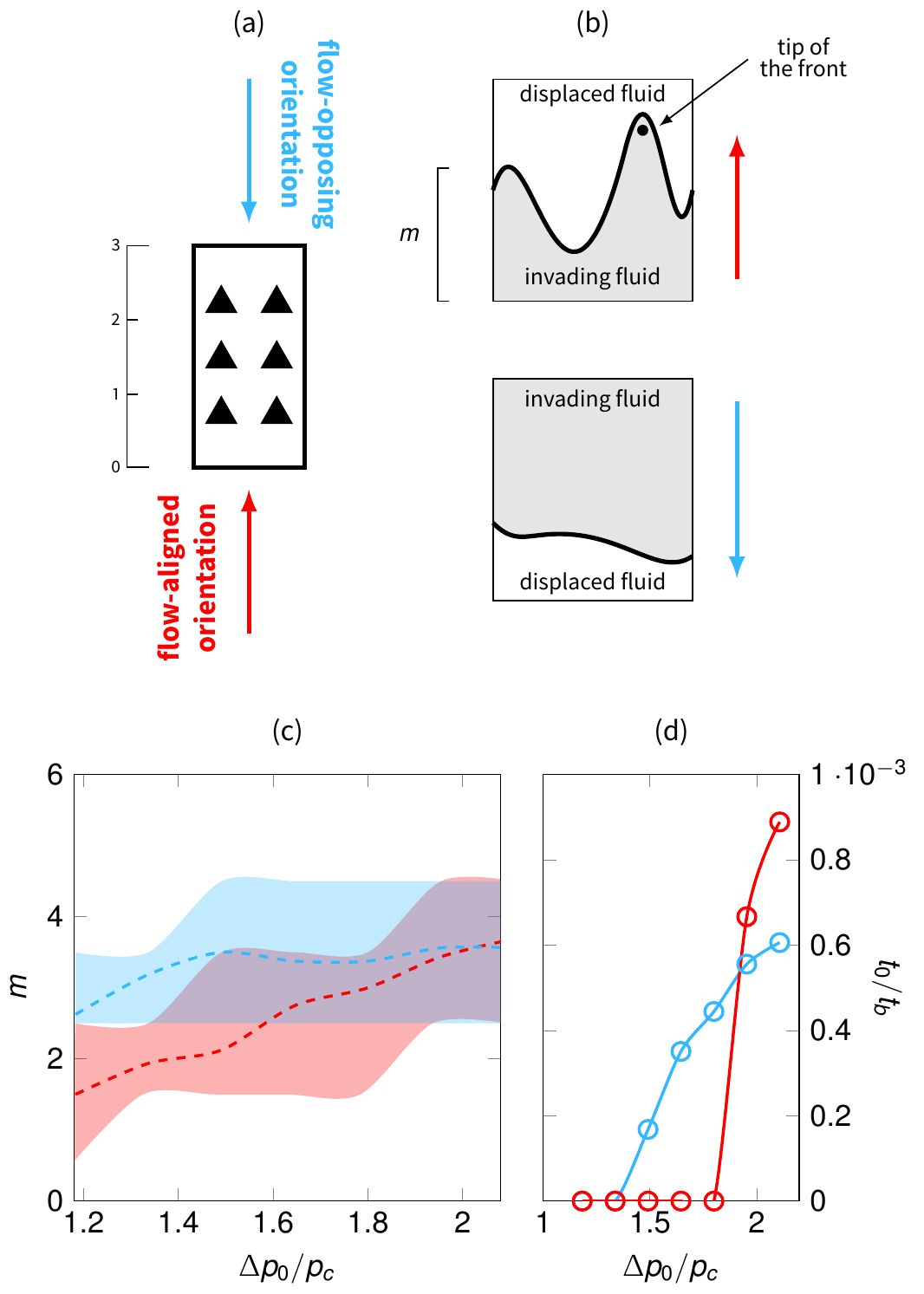}
\caption{(a) A sketch of the porous capillary valve exhibiting asymmetric invasion dynamics. The viscous phase (invading fluid) is injected  aligned (red) or opposing (blue) to the pillars, leading to a different front displacement, with the former case inducing capillary fingering and unstable displacement, see the sketch in (b). For certain values of the dimensionless forcing $\Delta p_0/p_c$, with $\Delta p_0$ the pressure drop and $p_c$ being the characteristic capillary threshold, the valve acts asymmetrically, i.e. conducting the flow primarily along one direction. The bottom panel (c) depicts the average front position $m$ (dashed lines) and the maximum and minimum penetration depth (the shaded area) at breakthrough time, computed in number of invaded pore rows, for aligned (red) and opposing (blue) injections. As the forcing increases, the medium becomes penetrable in both directions and the front stands at a maximum position $\max(m)>4$ which denotes that the viscous phase has somewhere invaded the fourth row and reached the outlet. In the last panel (d) the characteristic invasion rate $t_0/t_b$ is plotted for the different cases, with $t_b$ and $t_0=\mu_1/\Delta p_0$ being the breakthrough and characteristic times, respectively.}
\label{fig1}
\end{figure} 

We build up a porous medium that acts as a capillary valve: a system of flow control for which the magnitude of the flow resistance depends on the specific direction of injection of the viscous phase and on the driving force applied to the fluid. Triangular pillars compose the structure of the porous medium. 
We distinguish two configurations: when the viscous fluid is injected along the direction oriented with the triangular pillars, we refer the pillars as {\em flow-aligned}, in the sense that their cross section diminishes along the flow direction and that, during invasion, the viscous phase encounters firstly the triangle base and then its apex. In the opposite case, the viscous fluid is injected 
along the opposite direction
we refer the pillars as {\em flow-opposing} oriented, with the fluid firstly encountering the apex of the triangle. In the former case, the pore throat along the direction of injection presents a sudden constriction and a gradual enlargement. In the latter case, the pore throat size is gradually reducing and then a sudden enlargement is encountered. 

 The conceptualisation of this system is sketched on Fig.~\ref{fig1}, panels (a) and (b): when the flow is directed opposing to the pillars, the two-phase flow structure exhibits a rather compact uniform front and the drainage dynamics is slow but continuous. On the contrary, when the flow is directed aligned to the pillars the flow structure follows a dynamic roughening, with the formations of ramified fingers, and the drainage dynamics degrades rapidly, till a stable situation of equilibrium is reached. In the former case the viscous phase is often able to reach the outlet whereas, in the latter case, it can remain trapped in the medium, unable to reach the outlet. 

The panels (c) and (d) in Fig.~\ref{fig1} depict the asymmetric fluid behaviour observed in the performed numerical simulations. 
In the panel (c), we report the average front position at the time instant where the viscous fluid reaches the outlet or it has reached equilibrium, as a function of the driving force. The average front position $m$ is expressed in terms of units of invaded pore rows, where the total number of pore rows in this case is $M_\parallel =4$ (see also Eq.~\eqref{eq:sat}). 
The driving force is expressed in terms of maximum hydraulic pressure, or pressure drop, achievable in the system $\Delta p_0= -\nabla_x p_0 \  L_0$, where $-\nabla_x p_0$ is the applied body force that mimics the effect of a pressure gradient (see Section~\ref{sec:num}), $L_0$ is the total domain length along the direction of injection, and $p_{c}=\sigma/r$ is the characteristic pore-scale capillary threshold (with $\sigma$ the surface tension and $r$ a characteristic radius of curvature of the two-phase interface, whose measurement will be discussed on Section~\ref{sec:micro}).
In Fig.~\ref{fig1} (c) the shaded areas indicates the maximum and minimum values of the front position along the direction of injection, so that when $\max(m)>4$, the viscous injected phase has reached the outlet. The instant corresponding to such a situation is addressed as breakthrough time $t_b$, whereas $t_0=\mu_1/\Delta p_0$ is a characteristic viscous time. Thus the ratio $t_0/t_b$ represents a measure of the invasion rate and it is inversely related to the retention of the medium (i.e. the average amount of time that the viscous phase spends in the medium). The invasion rate $t_0/t_b$ is reported with varying the dimensionless driving force $\Delta p_0/p_c$ in Fig.~\ref{fig1} (d).

We observe a twofold functioning of the medium. The magnitude of the force driving the fluid determines the intensity of the flow, which significantly differs along the two directions of injection. The microstructural design of the porous system allows it to function as a valve that control the flow along specific directions, and, more specifically, as a system that conducts flows primarily in one direction. We can address such a peculiar functioning as {\em asymmetric flow conductance}, following the analogy with electronic components, and adopting a well-known terminology in hydraulics engineering
to define a directional-dependent permeability of the medium (i.e. the ratio between flow rate and pressure drop)~\cite{berg2012re}.
Three flow regimes can be distinguished from Fig.~\ref{fig1} (c) and (d): (i) for very low values of the fluid driving force the flow is impeded along both directions; (ii) as we increase the forcing the porous medium acts asymmetrically and a complete invasion is allowed only with flow-opposing orientation; (iii) for high values of the driving force, the difference between the two flow behaviours is reduced and the fluid flows through the medium along both directions (eventually with a higher transport rate with flow-aligned pillars). Such a behaviour unveils the possibility of regulating not only the allowed direction of the flow, but also the directional-dependent retention.

\begin{figure}
\includegraphics[width=0.99\linewidth]{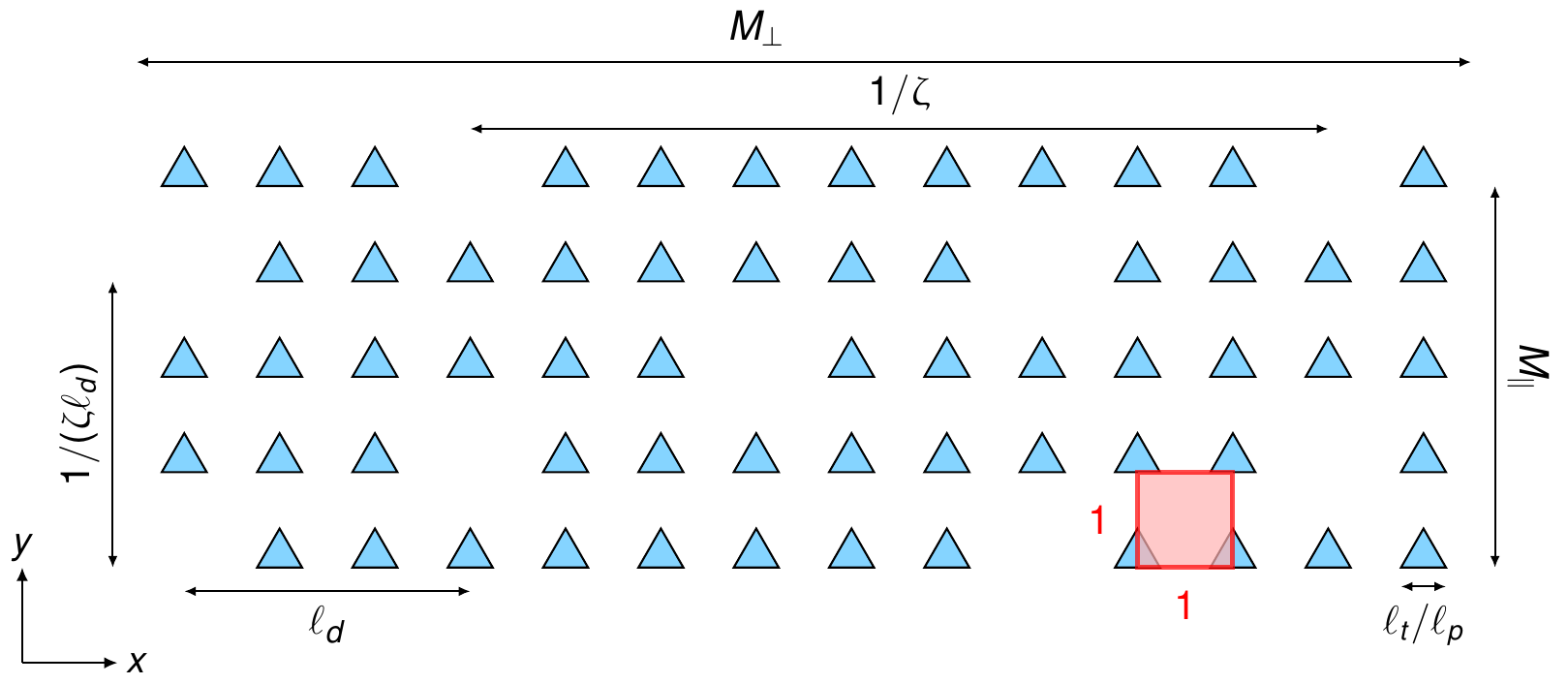}
\caption{
A sketch of the crystal-like anisotropic porous medium composed of triangular pillars with regularly distributed defects. All the characteristic lengths are given in dimensionless form, as functions of the characteristic pore size $\ell_p$. The porous domain is composed of $M=M_\parallel \ M_\perp$ pores. The representative elementary volume (REV) has the following length scales: the transversal length $1/\zeta$, the transversal distance between defects on contiguous rows $\ell_d$, and the periodicity of the system along the streamwise direction $1/(\zeta \ell_d)$. The characteristic pore throat size and the pillars size are $2r/\ell_p$ and $\ell_t/\ell_p$, respectively. Note that the same geometrical lengths and sizes are used for characterising the random configurations investigated in Section~\ref{sec:2p}, an example of which is given in Fig.~\ref{fig3} (a).}\label{fig2}
\end{figure}

We stress that the system is not subjected to gravity, the medium is neutrally wetted (contact angle $= 90^\circ$) and the observed asymmetric behaviour is purely induced by the microstructural design of the porous medium. The porous capillary valve design is based on two simple geometric principles: (i) to introduce anisotropy in the pore shape by changing its morphology and (ii) to tailor the pore-scale heterogeneities with the insertion of defects. The schematic Fig.~\ref{fig2} illustrates the involved geometrical parameters. The representative elementary volume is made up of equilateral triangle-shaped rigid solid pillars. By choosing such shapes we introduce an element of anisotropy at the pore scale.
Let us indicate with the symbols $i=\parallel,\perp,$ the directions connecting two pores separated by a pore throat along the streamwise $x$ and transverse $y$ directions, respectively. During an invasion event the viscous phase passing through the throat will exhibit a curved interface, whose radius of curvature is $r_i$ (see e.g. Fig.~\ref{fig7}, which we will discuss later in the paper). Then, we note that, in such a porous medium, the spatial distribution of capillary thresholds surrounding a pore $p_{c,i}=\sigma/r_i$ (i.e. the minimum capillary pressures required to invade the pores accessible through the contiguous pore throats) may be not isotropic.

For obtaining the capillary valve functioning, we choose a crystal-like configuration of the porous medium (we will also investigate random configurations later in the manuscript): the pillars are placed regularly at a distance $\ell_p$ in both $x$ and $y$ direction. The defects are identified by specific positions where such pillars are missing. We choose a percentage of defects $\zeta$ over the total number of pillars $N$: such defects are distributed regularly, in a staircase configuration, with $\ell_d$ the minimum transverse distance between defects of two contiguous rows (measured in number of pores or, equivalently, in number of distances $\ell_p$). The elementary representative volume (REV) is thus represented by a dimensionless  transversal and a longitudinal characteristic lengths $1/\zeta$ and $1/(\zeta \ell_d)$, respectively, as depicted in Fig.~\ref{fig2}.

With this geometrical configuration we are able to reproduce the valve behaviour depicted in Fig.~\ref{fig1}. The intuitive reason lies in the anisotropic and heterogeneous displacement of capillary thresholds provided by such a configuration. In the next sections we will present a numerical analysis able to explain the anisotropic and heterogeneous effects induced by the tuning of pore morphology and heterogeneity in such porous media.

\section{Numerical Methodology\label{sec:num}}

We make use of the Lattice Boltzmann methodology to simulate the two-phase flow. The simple form of the numerical algorithm based on this methodology allows us to deal with a great number of statistical realisations at a reasonable computational cost. The methodology accurately models the pore-scale two-phase invasion mechanisms given its ability to represent surface tension forces. The streaming-collision computation is performed according to:
\begin{eqnarray}
f_\xi(\mathbf{x}+\mathbf{c}_\xi\delta t, t+\delta t)-f_\xi(\mathbf{x},t) =  \nonumber \\ 
-\frac{\delta t}{\tau}(f_\xi(\mathbf{x},t)-f_\xi^{eq}(\rho,\mathbf{u}))+ F_\xi \delta t \ , \label{lbm1}
\end{eqnarray}
where $\xi$ labels the lattice direction in the D2Q9 lattice that we use, $\mathbf{x}=(x,y)$ is the position vector, $t$ is the simulation time, $\tau$ the relaxation time, $\mathbf{c}_{\xi}$ the discrete speed and $\delta t=1$ the simulation time step. The kinematic viscosity reads as $\nu=c_s^2 (\tau-0.5)$, with $c_s$ the speed of sound ($c_s^2=1/3$ for our lattice) We make use of the formulation based on a shift of the equilibrium velocity, i.e. $\rho \mathbf{u}_{eq} = \rho \mathbf{u} + (\tau - 1/2) F_\rho$, to define the equilibrium distribution function $f_\xi^{eq}$, as indicated in~\cite{chen2014critical}, with $F_\rho$ the Shan-Chen intermolecular force. Following the procedure described in~\cite{guo2002discrete}, with the forcing term in Eq.~\eqref{lbm1},  $F_\xi \propto -\nabla_x p_0$, we mimic a constant pressure gradient through the application of a body force acting along the streamwise direction, $ -\nabla_x p_0=\Delta p/L$. To represent surface tension forces, we instead make use of the Shan-Chen approach~\cite{shan1993lattice}, which mimics the intermolecular interaction through the computation of the density-dependent pseudo-potential function $\Psi(\rho)=1-e^{-\rho}$ and the effective intermolecular force $F_\rho$:
\begin{equation}
F_\rho(\mathbf{x},t)=-G\Psi(\mathbf{x},t) \sum\limits_\xi w_\xi \Psi(\mathbf{x}+\mathbf{c}_\xi,t)\mathbf{c}_\xi \ ,
\end{equation}
where $w_\xi$ represents the lattice weighting coefficient along the $\xi$-th direction. By choosing an appropriate value of the parameter $G=-5.5$ the interaction strength is sufficient to allow the separation of phases described by the non-ideal equation of state $P(\rho)=\rho c_{s}^{2}+G/2 \ c_{s}^{2}\Psi(\rho)^{2}$. At the fluid-solid boundaries we follow~\citep{de2011new} to set the equilibrium contact angle as $\theta=90^\circ$.
The fluid density and momentum are  determined via statistical averaging of the distribution functions as:
\begin{eqnarray}
\rho &=&\sum\limits_\xi f_\xi(\mathbf{x},t) \ , \\
\rho\mathbf{u} &=& \sum\limits_\xi f_\xi(\mathbf{x},t)\mathbf{c}_\xi+\frac{\delta t}{2}\frac{\Delta p}{L} + \frac{1}{2} F_\rho \ .
\label{lbm3}
\end{eqnarray}
For further details regarding the methodology, the specific algorithm and a series of validation test cases the reader is referred to~\citep{lbdm,maggiolo2019self,pettersson2020impact}.

\section{Two-phase flow simulations~\label{sec:2p}}

\subsection{Simulations set-up}

We perform numerical simulations of pore invasion dynamics during the injection of a viscous fluid within a medium initially filled with a less viscous one, so that the dynamic viscosity ratio is $\mu_1/\mu_2 = 35 \gg 1$. To better explain and understand the meaning of the results presented in Section~\ref{sec:valve},
we investigate different geometrical configurations, represented by the characteristic length scales sketched in Fig.~\ref{fig2}. We firstly investigate three random configurations, characterised by a fraction of defects $\zeta=0.03,0.06$ and $0.10$. In these three configurations, the defects are introduced by randomly placing the percentage $\zeta$ of triangular pillars according to a uniform distribution. The other $[(1-\zeta)N]$ pillars are regularly placed at a constant distance $\ell_p$. See Fig.~\ref{fig3} (a) for an example of such random configuration. To increase the statistical accuracy of our data, for each of the three random configurations we perform numerical simulations in 4 different random realisations of the geometry.

To the characteristic pore length $\ell_p$ and characteristic size of the triangular pillar $\ell_t$, correspond 28.6 and 13.4 computational discretised elements, respectively. We found in these values a good compromise between numerical accuracy and computational cost required to perform several simulations carrying sufficient statistical information. 
The typical simulation set-up comprises an area of the viscous fluid 1, whose length and width are  $L$ and $H$, respectively, placed just above or below a porous medium of the same size. The porous medium and the viscous fluid area are contained within a bi-periodic domain of length $L_0/\ell_p=14.7 > 2 L /\ell_p$. Periodic boundary conditions are imposed along the streamwise, $x$, and transverse, $y$, directions. After the initial time instant the viscous phase is injected into the medium under the action of the body force $- \nabla_x p_0$ acting along the positive (flow aligned pillars) or negative (flow-opposing oriented pillars) $x$ direction.

We also perform numerical simulations in three crystal-like structures, where the solid pillars composing the medium are regularly distributed in the lattice and the defects are introduced by simply removing the pillars at specific locations. The  last crystal-like configuration is the one that provides the valve behaviour sketched in Fig.~\ref{fig1}. The comparison between the first (random) and second (crystal-like) set of simulations allows us to quantify the difference between a stochastic and a fully deterministic geometrical configurations. A list of the geometrical lengths and parameters characterising the geometries is provided in Tab.~\ref{tab1}.

\begin{table}
 \caption{\label{tab1}The geometrical parameters that characterise the investigated cases. The dimensions of the medium are $L$ and $H$ that, computed in terms of number of pores, correspond to $M_\parallel=L/\ell_p$ and $M_\perp=H/\ell_p$ along the longitudinal and transverse directions, respectively. The last crystal-like structure listed in the table and denoted with an asterisk is the one that provides the capillary valve functioning described in Section~\ref{sec:valve}.}
\begin{ruledtabular}
\begin{tabular}{l c c c c c c}
Case & $\zeta$ & $1/\zeta$ & $\ell_d$ & $1/(\zeta\ell_d)$ & $M_\parallel$  & $M_\perp$\\
\hline 
Random & 0.03 & 33 &  9.3 & 3.6 & 6 & 42\\
Random & 0.06 & 16 &  7.6 & 2.2 & 6 & 42\\
Random & 0.10 & 10 &  6.9 & 1.5 & 6 & 42\\
Crystal-like & 0.03 & 32 & 16 & 2 & 6 & 32\\
Crystal-like & 0.06 & 16 & 8 & 2 & 6 & 16\\
Crystal-like$^*$	& 0.06 & 16 & 4 & 4 & 4 & 16\\
\end{tabular}
\end{ruledtabular}
\end{table}

\subsection{Pore invasion dynamics and kinetic roughening}

\begin{figure*}
\includegraphics[width=0.79\linewidth]{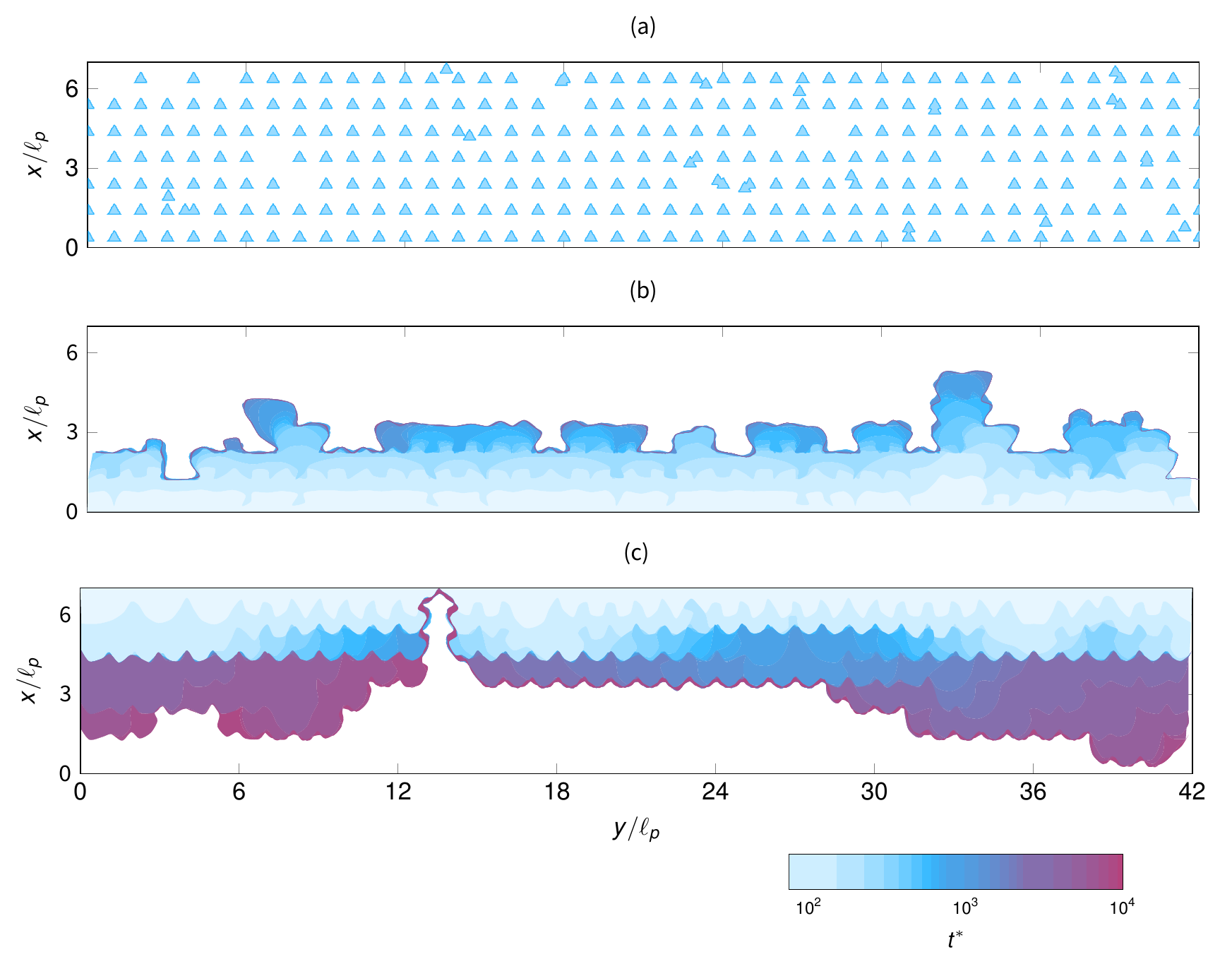}
\caption{
An example of a random configuration of the anisotropic medium. The investigated system consists of a fluid invading a porous medium composed of $N$ solid pillars initially filled with a less viscous fluid. The pillars composing the medium are $(1-\zeta)N$ equilatelar triangles uniformly distributed with inter-pillar distance $\ell_p$ and $\zeta N$ randomly distributed. In the upper panel (a), a porous medium composed of triangular-shaped pillars with a percentage of defects $\zeta=0.06$ randomly distributed is depicted. In the lower panels, the dynamics of the invading front with flow-aligned (b) and flow-opposing oriented (c) pillars is presented. The shading colours indicate the position of the interface at different invasion times $t^*=t/t_c$. It is interesting to notice that, when the pillars are flow-opposing oriented (c), successive invasion events often occurs along a serpentine-wise direction, with a streamwise invasion when the front reaches a defect and jumps to the successive pore row, followed by many transversal invasion events.}\label{fig3}
\end{figure*}

We firstly focus our attention on the pore invasion dynamics occurring in the random samples. In Fig.~\ref{fig3} the front position at different dimensionless time $t^*=t/t_0$ is depicted, for the case $\zeta=0.06$, where $t_0=\mu_1/\Delta p_0$ is a characteristic time of our system. We anticipate here that $\Delta p_0=-\nabla_x p_0\ L_0$ is the maximum pressure drop achievable in the system or, equivalently, the maximum hydraulic head. We further discuss later the significance of this quantity.

In the investigated two-phase system, two competing forces are determining the fluid displacement and the pore invasion events. We expect viscous forces contributing to stabilise the front. On the other hand, the geometrical microstructure can counteract such a stabilising mechanism and, under certain conditions, promote unstable displacement and capillary fingering phenomena.
We observe from Fig.~\ref{fig3} that indeed these two mechanisms are both playing a role in determining the front configuration. We note that the presence of defects affects the spatial distribution of the viscous phase; indeed defects provide sites with higher probability of invasion, being them connected with a higher number of contiguous pores.
We also immediately notice that, depending on the medium orientation relative to the flow direction, i.e. with flow-aligned or flow-opposing oriented pillars, the two-phase interface structure changes, with the former case being characterised by a rough front with fingers of the size of few pores. In such a case we intuitively recognise a form of instability that promotes kinetic roughening and dominates over the stabilising viscous forces for a length comparable with few pores. We will come back later in the next section to discuss this competitions between stabilising and destabilising mechanisms.

\begin{figure*}
\includegraphics[width=0.89\linewidth]{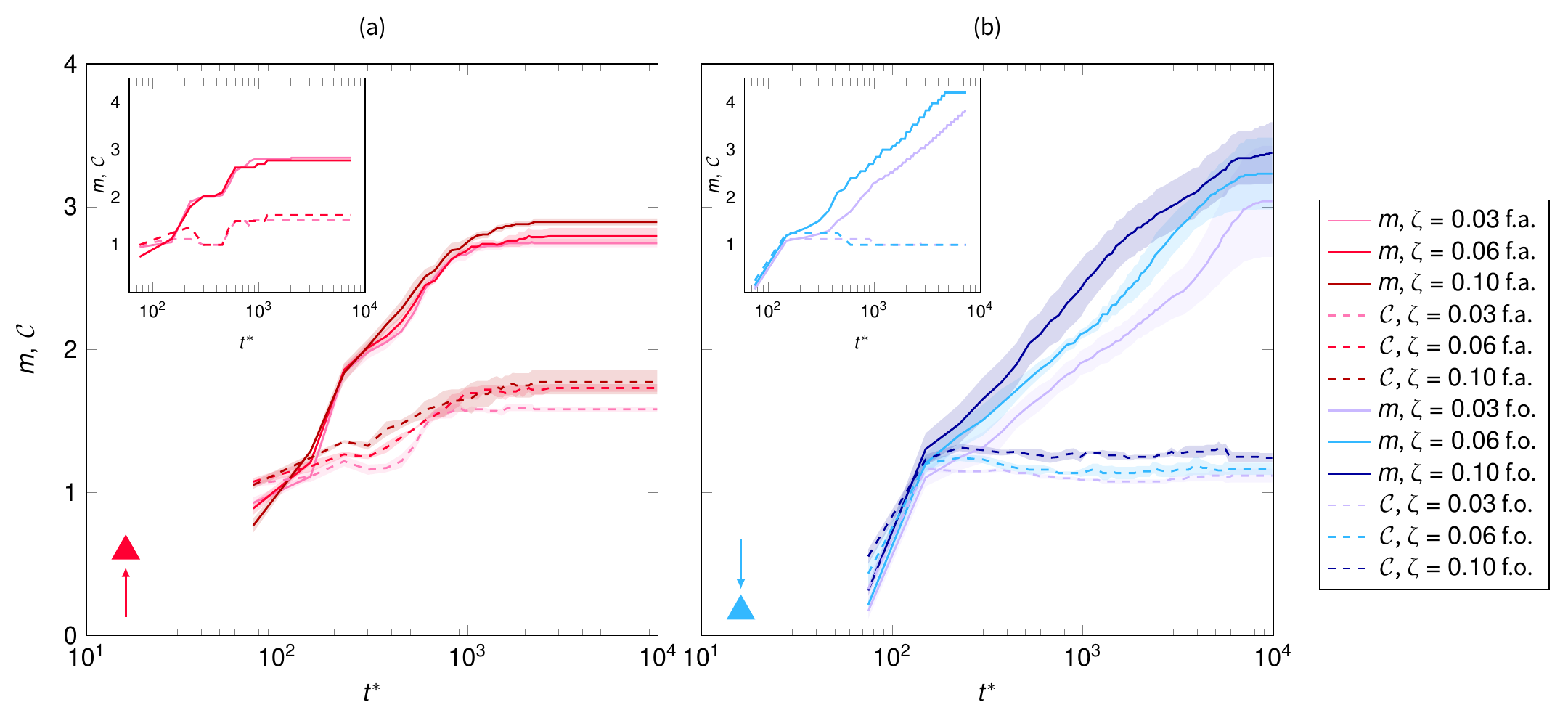}
\caption{Dynamics of the cumulative number of pore invaded over the average number of pores per row, $m(t)$ (solid lines), and number of pore throats at the liquid-gas interface over the average number of pores per row, $\mathcal{C}(t)$ (dashed lines). The dynamics is represented for flow-aligned  (a) and flow-opposing (b) oriented triangular pillars. In the insets, the cases with regularly placed defects are shown (crystal-like structures). The solid lines refer to the average value between different statistical realisations, the shaded areas represent the statistical uncertainty, estimated via the standard error (the standard deviation divided by the square root of the number of realisation). The color gradient of the solid lines is related to different fractions of defects $\zeta$, with darker lines denoting a higher fraction. We notice that when the pillars are flow-opposing oriented, the roughness of the front, $\mathcal{C}$, is reduced while the number of pore invaded (saturation), $m(t)$, is enhanced at long times, for both the random and crystal-like configurations.}
\label{fig4}
\end{figure*}

Following the aforementioned observation, we compute the dynamics of two important quantities, the number of pores invaded and occupied by the viscous fluid $s(t)$ and the number of pore throats lying along the two-phase interface $c(t)$, at a time instant $t$:
\begin{eqnarray}
m(t) = \frac{s(t)}{M_\perp} \ , \label{eq:sat} \\
\mathcal{C}(t) =\frac{c(t)}{M_\perp} \ . \label{eq:rough} 
\end{eqnarray}
These two quantities are non-dimensionalised with the average number of pores per row $M_\perp=M/M_\parallel$, where $M$ is the total number of pores identifiable in each configuration and $M_\parallel=L/\ell_p$ is the thickness of the medium measured in units of pore size (see Tab.~\ref{tab1}). The former quantity, $m(t)$, is a measure of the saturation of the porous medium whereas the latter, $\mathcal{C}(t)$, quantifies the roughness of the interface. 

From Fig.~\ref{fig4} we can confirm the qualitative observation we have previously pointed out.
The case with flow-aligned pillars is shown on the left panel of Fig.~\ref{fig4}: the medium saturation initially increases rapidly in time and then slows down significantly after less than three rows are fully invaded, i.e. $m< 3$. At the same time, the roughness of the surface follows a similar dynamics and eventually reaches $\mathcal{C}\sim 2$. On the other hand, when the pillars are flow-opposing oriented (downward injection, right panels), the medium saturation, $m(t)$,  increases in a more continuous manner, eventually reaching $m> 3$, and the roughness after the initial stage is stabilised at a value $\mathcal{C}(t)\sim 1$. The latter value of the dimensionless roughness indicates that the number of pores lying at the two phase interface approximately equals the average number of pores per row, i.e. $c(t)\sim M_\perp$, and thus that the two-phase front is flat, as we can intuitively observe also from Fig.~\ref{fig3}. 

We recognise that for a lower value of the front roughness (flow-opposing oriented pillars), a higher number of invasion events occurs at long times and the porous media is more easily filled. On the contrary, an excessive increase of the front roughness appears to slow down the invasion process. We observe these two opposite behaviours for all the random cases investigated and for the crystal-like structures (insets on Fig.~\ref{fig4}), without noticing any substantial and quantitative difference. This consideration suggest us a very important insight: in such conditions, the effects of the anisotropy of the pore shape (pore morphology) are dominant whereas the amount of defects and their spatial random arrangement (pore-scale heterogeneity) have a small effect on the front roughness and temporal fluid-fluid displacement.

\subsection{Capillary pressure distribution and anisotropic unstable two-phase displacement}

The capillary pressure distribution at the two-phase interface are important observables to analyse the invasion dynamics. Since the invading fluid is more viscous than the invaded one, the stabilising effect of viscous forces should  induce a pressure drop along the {\em invading fluid structure}, e.g. along an invading finger~\cite{cottin2010drainage}. We recognise that defects are a more probable source of invasion events since, on average, they are connected with a higher number of adjacent not invaded pores. We can thus conjecture that a defect represents the source site for a cascade of successive invasion events, i.e. the base of an invading fluid structure or, equivalently, the pore site that separates the invasion process of a pocket of connected pores with similar morphological properties. 

When the two-phase interface crosses a pore throat to rapidly invade a large pore site, such as a defect, the pressure measurement of the invading fluid shows a local minimum, a phenomenon usually referred as a {\em burst event}~\cite{roux1989temporal}.
After this process, the pressure at the invaded defect experiences an increase until it overcomes again the capillary threshold for invading a neighbouring pore. We can thus define the {\em invading fluid structure} as the dynamic invasion process of pore sites occurring between burst events. 

\begin{figure}
\includegraphics[width=0.99\linewidth]{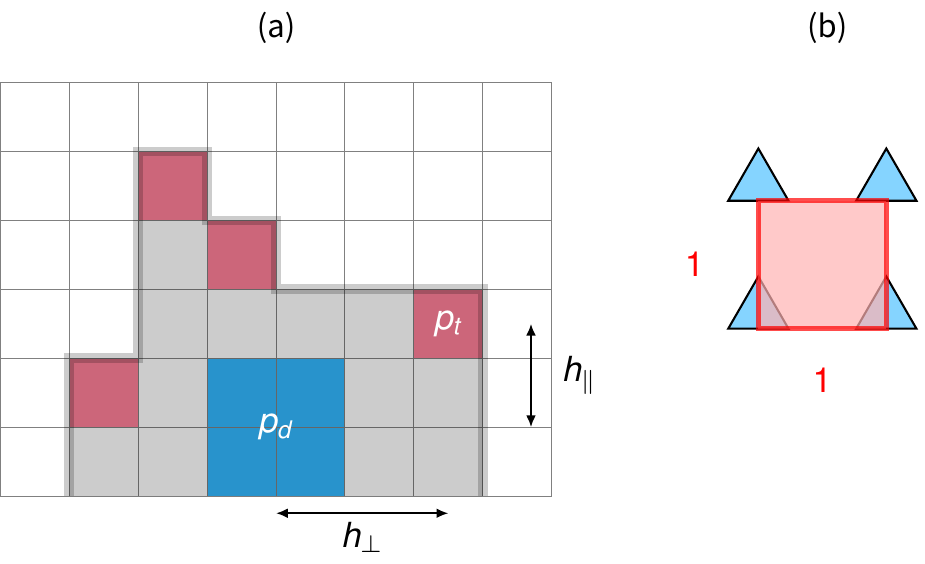}
\caption{\label{fig5} Sketch of an invading fluid structure after a burst event (the full invasion of a defect). The base of the invading fluid structure is located in correspondence of the defect (blue square). Each square of the sketch in the panel (a) corresponds to a unitary pore space composing the medium (b). The tips of the front are defined as the invaded pore at the two-phase interface that are connected with at least two not invaded pores (red squares). To each tip, the closest defect is identified and the Cartesian components of the base-tip distance $h_\parallel$ and $h_\perp$ are measured and used for the computation of the dimensionless pressure gradients defined in Eq.~\eqref{eq:cap}.}
\end{figure}

To characterise such an invasion process, we measure the pressure at the tip and base of the invading fluid structure. A tip is defined as the pore site invaded by the viscous fluid, at a given time instant, that is connected with a relative higher number of not invaded pores (growth sites). As a consequence, in such a pore, the probability for successive invasion events should be relatively higher compared to other pore sites connected with a lower number of growth sites. We choose to define the tips as the pores with a minimum of two neighbouring growth sites.
After a tip is identified, we search for the closest invaded defect (base). The base-tip distance is identified by $h=(h_\parallel^2+h_\perp^2)^{1/2}$, where with $h_i$ and the subscript $i=\parallel,\perp$ we label the streamwise and transverse Cartesian components of such a distance, respectively.  A schematic of the method used for the identification of the base and tip pore sites is presented in Fig.~\ref{fig5}.

During the invasion process of a fluid structure, we expect the pressure at the defect, $p_d$, to be higher than the one computed at the tip, $p_t$. Indeed, the invading fluid moves within the two-dimensional porous structure, viscous forces acts between the fluid and solid phases and a pressure gradient between base and tips is established. Such a pressure gradient quantifies the velocity of invasion events along the invading fluid structure defined between the base and the tips. It also conveys information about the distribution of capillary pressures. Since the viscous pressure drop on the invaded phase is much smaller ($\mu_2\ll \mu_1$), we can consider its pressure $p_2$ rather constant and thus we can have an approximate measure of the difference between capillary pressures in the vicinity of the base and tips of such an invading fluid structure, $\Delta p_{cf}$, through the computation of $p_d-p_t$. The dimensionless pressure gradients of the invading fluid structure, along the streamwise and transverse direction, are defined in the two-dimensional space as:
\begin{equation}
-\nabla_i^* p^*=\frac{p_d^*-p_t^*}{|h^*_i|}  \sim \frac{ \Delta p_{cf} }{|h_i|}\frac{\ell_p}{p_c} \ , \label{eq:cap}
\end{equation}
where the dimensionless pressures are $p_d^*=p_d/p_c$ and $p_t^*=p_t/p_c$.
The pressure difference $p_d^*-p_t^*$ in Eq.~\eqref{eq:cap} is divided by the dimensionless absolute value of the Cartesian components of the base-tip distance $h_i^*=h_i/\ell_p$, and with the symbols $i=\parallel,\perp$ we label the streamwise and transverse directions.

\begin{figure}
\includegraphics[width=0.79\linewidth]{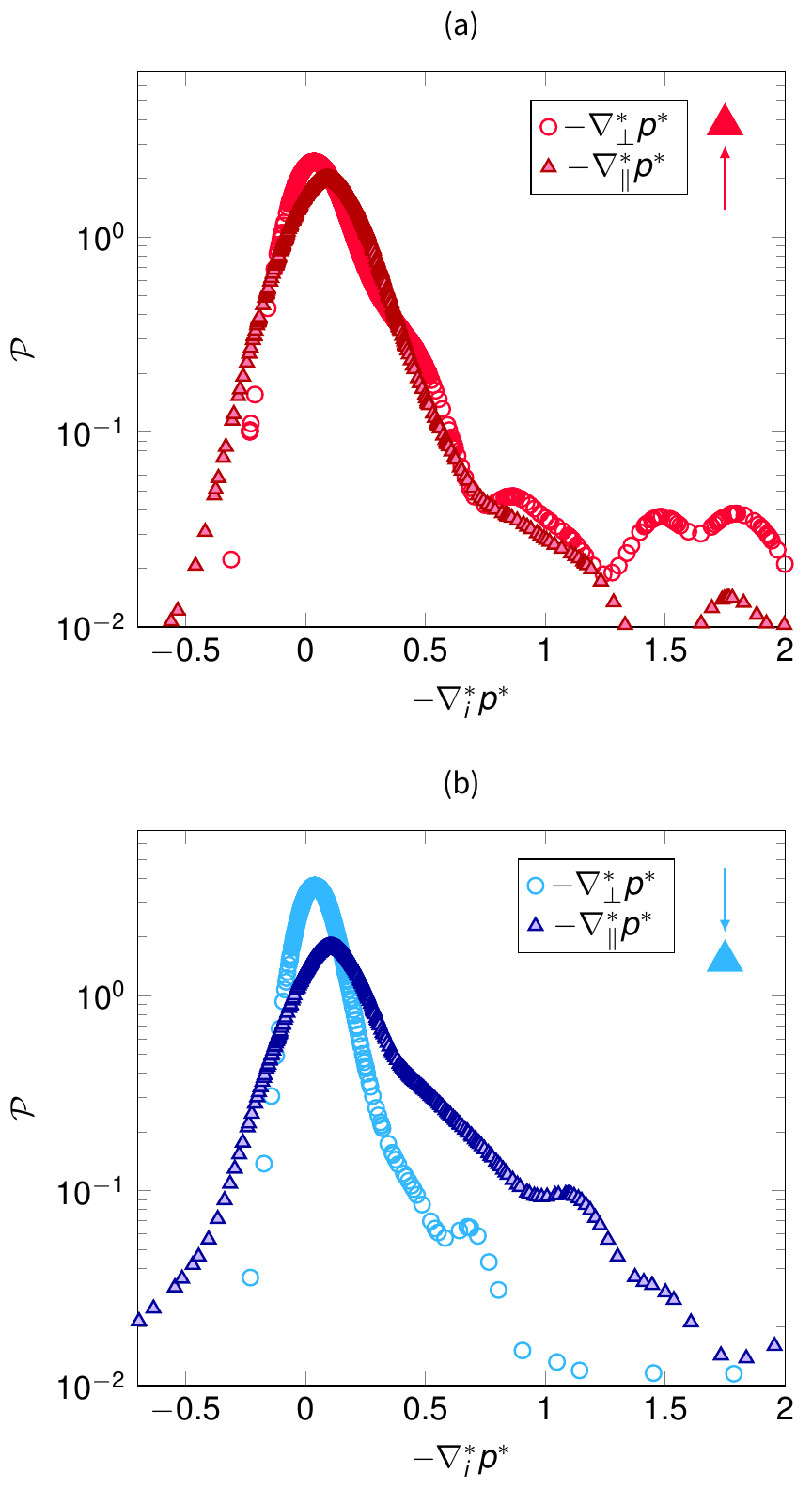}
\caption{\label{fig6}
Probability distribution functions $\mathcal{P}$ of the pressure gradients established between bases and tips of the invading fluid structures, $-\nabla^*_i p^*$, as defined in Eq.~\eqref{eq:cap}, with $i=\parallel,\perp$ indicating a pressure gradient along the streamwise and transverse directions, respectively. We report the case of flow-aligned (a) and flow-opposing oriented (b) triangular pillars with random defect distributions. The probabilities are computed at the simulation times $10^2<t^*<10^4$. We observe a substantial difference between streamwise and transverse pressure gradients for the case with flow-opposing oriented pillars (b).}
\end{figure}

In Fig.~\ref{fig6} the probability distribution functions of the fluid structure pressure gradients are showed. The probabilities refer to all time instants and thus they well depict  the global functioning of the invasion dynamics as a function of the spatial capillary pressure distributions.
The occurrence of a low value of $-\nabla^*_i p^* $ implies that the tip and base of the invading front have a very similar value of pressure, despite the relative long distance that separates them. In an equivalent perspective, the viscous pressure drop between base and tip of the front is relatively low for that distance. Such a low value of  $-\nabla^*_i p^*$ thus corresponds to an unstable two-phase displacement, because it suggests that the capillary pressures at the pores in the vicinity of the base and in the vicinity of the tip are similar, as well as their percolation probability for successive invasion events. 

On the other hand, the occurrence of stable two-phase fluid displacement is instead indicated by high values of  $-\nabla^*_i p^*$, which suggest a significant pressure drop between bases and tips for relatively short distances. As a consequence the capillary pressure at the tip of the fluid structure is much lower than that at the base, implying a lower probability of successive invasion events at the front of the invading  fluid structure. 

The probability distribution functions showed in  Fig.~\ref{fig6} exhibit a rather similar shape.   
They present a peak in proximity of zero, an indication of a significant amount of unstable displacements, a rapid decay for negative values, which possibly addresses the rare occurrence of uncorrelated pressures, and a right heavy tail for larger positive values, which depicts the presence of a wide distribution of fluid structures stabilised by viscous forces. The striking difference between the distributions is found when comparing the two cases with different medium orientation relative to the direction of injection. When the triangular pillars composing the medium are flow-aligned, Fig.~\ref{fig6} panel (a), the distributions of base-tip pressure gradients are similar, irrespective of the chosen Cartesian components, $h_\parallel$ or $h_\perp$, through which they are computed. This observation indicates that fluid invasion events occurs similarly in the streamwise and transverse direction and that the probability of stable-unstable events is evenly distributed in the two-dimensional space.

On the contrary, when the triangular pillars composing the medium are flow-opposing oriented,  Fig.~\ref{fig6} panel (b), we observe that  along the transverse direction the base-tip pressure gradients decays more rapidly compared to the streamwise direction. Such a striking difference points to an anisotropic two-phase flow behaviour, where a less stable mechanism of invasion occurs along the transverse direction.

\subsection{Anisotropic distribution of capillary thresholds~\label{sec:micro}}

The asymmetry and anisotropy of the invasion process in the porous medium is the key to interpret the observed differences in the growth of the interface roughness and average front position, when changing the medium orientation with respect to the direction of injection. Such a characteristic can also unveil the possible design of capillary-valve porous media, as the one described at the beginning of the present manuscript.

\begin{figure}
\includegraphics[width=0.59\linewidth]{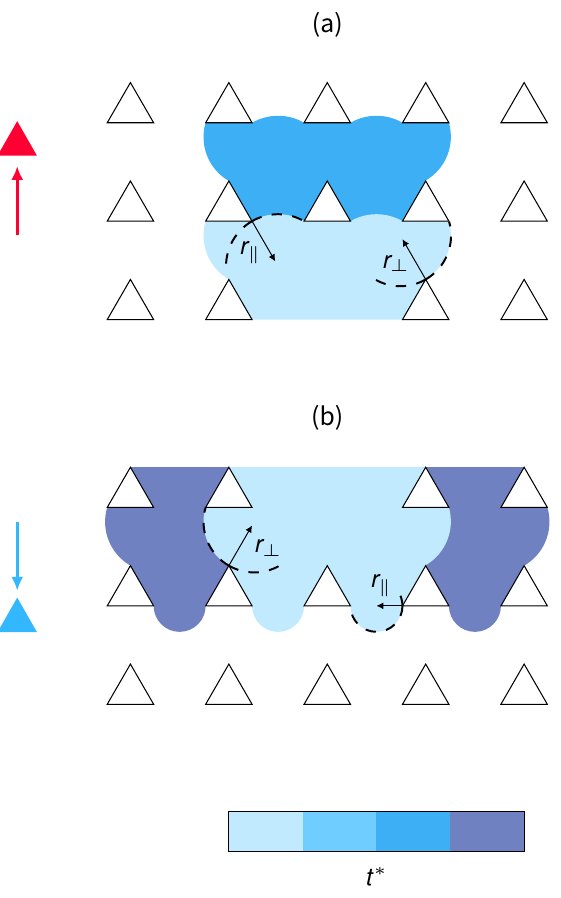}
\caption{
Pore invasion events depends on the isotropic/anisotropic distribution of capillary thresholds. In the case of flow-opposing oriented pillars (b), a marked anisotropic pore invasion mechanism is observed. The capillary threshold at the pore throat connecting two pores longitudinally or transversally is $p_{c,i}\sim \sigma/r_i$, with $i=\parallel,\perp$ indicating the streamwise and transverse directions, respectively. Here $r_i$ represents the minimum radius of curvature of the moving meniscus at the pore throat. Since the medium is neutrally wetted, the two-phase interface must form a  contact angle of 90$^\circ$ at the three-phase contact point while invading a pore throat. Using simple geometrical consideration we obtain for the flow-aligned pillars (a) $r_\parallel\sim r_\perp\sim \ell_t$, marking an isotropic spatial distribution of percolation threshold. On the other hand, for the flow-opposing pillars (b) it results instead $r_\parallel \sim r_\perp/2 \sim \ell_t/2$, which denotes a lower capillary threshold for invasion along the transverse direction.}\label{fig7}
\end{figure}

We have already intuitively observed that the peculiar morphology of the pore, induced by the presence of specifically oriented triangular pillars, affects the spatial distribution of capillary thresholds. Since in our simulations the solid surfaces are neutrally wetted (contact angle $90^\circ$), we can estimate the capillary threshold, i.e. the minimum pressure needed to invade an adjacent pore, on the basis of the radius of curvature of the meniscus when the viscous fluid crosses a pore throat:
\begin{equation}
p_{c,i}\sim \frac{\sigma}{r_i} \ ,
\end{equation}
where again $i=\parallel,\perp$ indicates the streamwise and transverse directions, respectively, according to the direction of invasion, and $r_i$ represents the minimum radius of curvature of the moving meniscus at the corresponding pore throat. Such an estimation lead us to compute an isotropic distribution of capillary thresholds when the pillars are flow-aligned, Fig.~\ref{fig7} panel (a). On the contrary, we must recognise that the distribution of capillary thresholds is markedly anisotropic in the case of flow-opposing oriented pillars, Fig.~\ref{fig7} panel (b). In particular, for such a case, the capillary threshold characterising transversal invasion events is much lower than that encountered along the streamwise direction.

As a consequence, for the case of flow-aligned pillars, we observe a rather isotropic invasion dynamics, with the formation of short fingers triggered by defects, together with a rough interface and an even spatial distribution of stable and unstable invasion events (see Fig.~\ref{fig6}). Albeit their amount affects only slightly the invasion dynamics (see Fig.~\ref{fig4}), defects are still important in inducing the observed fingering. When the pillars are flow-opposing oriented, we observe instead a marked and unstable displacement of fluid along the transverse direction and a less probable invasion along the streamwise one, which is limited by the high values of the capillary thresholds and viscous forces that contributes to stabilise the front. We are not surprised thus to also observe in such a case a flat front of the invading fluid and a reduced two-phase interface roughness.  

\begin{figure*}
\includegraphics[width=0.89\linewidth]{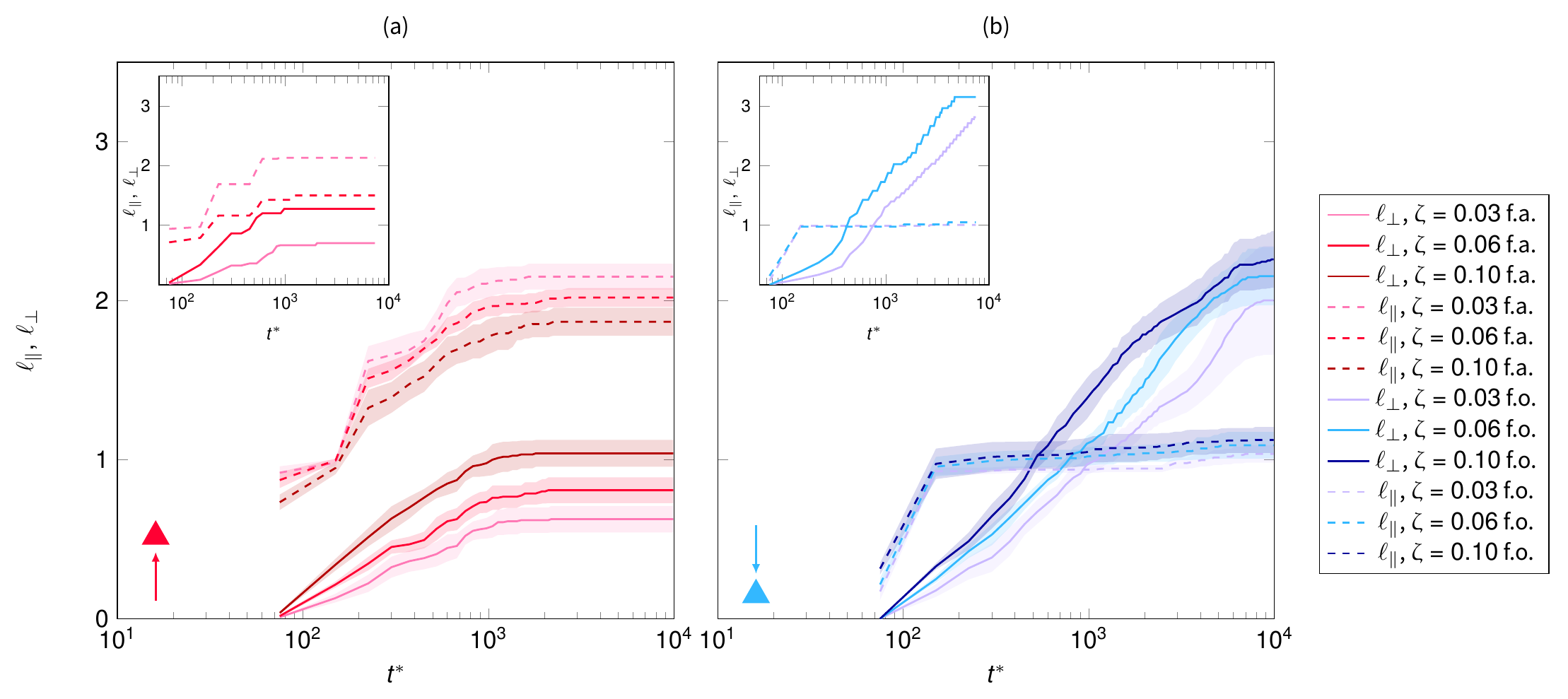}
\caption{Dynamics of the cumulative number of pore invaded transversally, $\ell_\perp(t)$ (solid lines), and along the streamwise direction, $\ell_\parallel (t)$ (dashed lines), over the  average number of pores per row $M_\perp$. The case of flow-aligned (a) and flow-opposing oriented (b) pillars are represented. In the insets, the cases with regularly placed defects are shown. We observe a similar invasion rate $\mathrm{d} \ell_i (t) /  \mathrm{d} t^*$, for $i=\parallel,\perp$ in the case of flow-aligned pillars (a), since along both directions the capillary thresholds are similar, i.e. $p_{c,i}=\sigma/\ell_t$, as depicted in Fig.~\ref{fig7}. On the contrary, when the pillars are flow-opposing oriented (b), the invasion rate is substantially higher along the transverse direction for which the capillary threshold is substantially lower, i.e. $p_{c,\perp}=\sigma/\ell_t$. In both cases, at the initial stage, for very short dimensionless times, the viscous fluid invades the first pore row along the streamwise direction, subjected to inertial forces.}
\label{fig8}
\end{figure*}

In order to further corroborate our observations, in Fig.~\ref{fig8} we calculate the number of invasion events occurring along the transverse and streamwise directions, for all the investigate cases. This calculation indeed confirm that, after the initial invasion stage, the dynamic rate of the invasion events occurs similarly along the streamwise  $\ell_\parallel$ and transverse  $\ell_\perp$ directions for the flow-aligned pillars, with a slightly more pronounced streamwise invasion triggered by defects. On the contrary the invasion occurs almost exclusively along the transverse direction for the flow-opposed pillars, marking a strong anisotropic and serpentine-wise fluid invasion mechanism (see e.g. Fig.~\ref{fig4}). We also observe that the large majority of the invasion events occur crossing pore throats characterised by the lower capillary threshold. Thus, the invasion dynamics is dominated by a characteristic capillary threshold $p_c=\sigma/r$, with $r\sim\ell_t$ (see Figs.~\ref{fig7} and \ref{fig8}). 

\subsection{The role of front roughness in determining the invasion dynamics ~\label{sec:rough}}

The anisotropic distribution of capillary thresholds triggers different invasion dynamics according to the medium orientation, resulting in substantially different invading fluid structures and configurations of the front roughness. We recognise that such different scenarios can lead to different rates of invasion, as we have seen in Fig.~\ref{fig4}, and possibly impede the full invasion of the medium in some cases. This reasoning suggests that the quantification of an effective resistance related to the front roughness, or, equivalently, an effective capillary pressure induced by the microstructure, can support the mathematical description of the two-phase flow transport equations in the porous medium. The quantification of the effective capillary pressure is indeed commonly used for closing the system of equations of two-phase flows in porous media~\citep{whitaker1986flow}. 

Here with $p_{c,\mathit{eff}}$ we thus indicate the effective capillary pressure contribution related to the curvature of the two-phase interface in the medium. To close the transport equation for the invading phase we need to draw up few assumptions: (i) we postulate that the transport equation of the invading phase can be described by a Darcy-like formulation where (ii) we approximate the pressure gradient driving the invading phase on the basis of the balance between effective pressure drop and viscous forces, which are proportional to the front penetration depth $m(t)$. In particular, we write the pressure gradient driving the invading fluid along the streamwise direction, averaged between each parallel pore column, as:
\begin{equation}
-\nabla_x  p_1(t)  \sim \frac{1}{m(t) \ell_p} \big (-p_{c,\mathit{eff}} (t) + \Delta p_0\big ) \ , \label{eq:nabla}
\end{equation}
where $\Delta p_0=-\nabla_x p_0\ L_0$ is the maximum hydraulic pressure, or pressure drop, achievable in the system, in each parallel pore column. In writing Eq.~\eqref{eq:nabla}, we have neglected the effects of the displaced fluid motion, since $\mu_2\ll \mu_1$. The meaning of Eq.~\eqref{eq:nabla} is to estimate the average pressure gradient acting on the invading fluid portion $m(t)\ell_p$ through the quantification of the maximum pressure drop reduced by the contribution of the macroscopic effective capillary pressure $p_{c,\mathit{eff}}$. In an different but equivalent perspective, Eq.~\eqref{eq:nabla} represents the energy budget available to the invading fluid, which is given by the difference between the total energy provided to the system and the energy partially spent to create the two-phase interface. In the limit case $p_{c,\mathit{eff}}=\Delta p_0$, the two-phase interface is blocked, because it cannot overcome the effective capillary threshold, the flow velocity is null, and the pressure exhibits the hydraulic distribution for a quiescent fluid. Following Eq.~\eqref{eq:nabla}, we can write a Darcy-like formulation for the invasion process:
\begin{equation}
\frac{\mathrm{d} m(t)}{\mathrm{d} t} = \frac{K^*}{m(t)} \frac{\Delta p_0}{\mu_1} \bigg (1 -\frac{p_{c,\mathit{eff}}(t)}{\Delta p_0} \bigg )  \ , \label{eq:darcy1}
\end{equation}
where $K^*=K/\ell_p^2$ is a characteristic dimensionless permeability and the product $K^* (1-p_{c,\mathit{eff}}/\Delta p_0)$ can be interpreted as an effective medium permeability or flow conductance. 

The exact quantification of $p_{c,\mathit{eff}}(t)$ is not trivial, since it depends on the local distribution of capillary pressures along the advancing two-phase interface. Let us assume that each invasion event occurs when the local microscopic capillary pressure overcomes the corresponding pore-scale capillary threshold. Since, as we have seen, the large majority of the invasion events occurs along pore constrictions characterised by $p_c=\sigma/\ell_t$, we can estimate the local capillary pressure at the advancing two-phase interface via $p_c$. The macroscopic capillary pressure $p_{c,\mathit{eff}}(t) M_\perp$ represents the sum of each capillary pressure contribution at the pore scale. We thus compute $p_{c,\mathit{eff}}(t) M_\perp \sim c(t)\ p_c$ and $p_{c,\mathit{eff}}(t)\sim \mathcal{C}(t)\ p_c$, where we remind that $\mathcal{C}(t)= c(t) / M_\perp$ is the fraction of pore throats belonging to the two-phase interface at a time instant $t$, with respect to the average number of pores in a row. 

Following the same reasoning, we define $\mathcal{C}_0$ as the maximum number of possible pore throats belonging to the two-phase interface, when the fluid is subjected to a pressure difference $\Delta p_0$. It follows that $\Delta p_0=\mathcal{C}_0 \ p_c$ and integrating and rewriting Eq.~\eqref{eq:darcy1} we obtain:
\begin{equation}
m(t)^2 = K^* \int_0^{t^*} 1 -\frac{\mathcal{C}(t)}{\mathcal{C}_0} \ \mathrm{d} t^* \ ,  \label{eq:darcy2}
\end{equation}
where $t^*=t/t_0$ and $t_0=\mu_1/\Delta p_0$. In the limit case $\mathcal{C}(t)=0$, Eq.~\eqref{eq:darcy2} provides the scaling $m(t)\propto {t^*}^{1/2}$, thus recovering Lucas-Washburn solution for forced imbibition ~\cite{hilpert2010explicit} and the solution of Richards equation for constant sorptivity~\cite{philip1969theory}. Equation~\eqref{eq:darcy2} states that invasion events occur as long as $\mathcal{C}(t) < \mathcal{C}_0$ and there is still available energy for sustaining them. As the roughness increases, such a budget diminishes reducing the invasion rate. The application of such a model, which makes use of the characteristic capillary threshold $p_c$, leads to $\mathcal{C}_0\sim \Delta p_0 /p_c \sim 2$ for the random and crystal-like configurations investigated in this Section~\ref{sec:2p}. We eventually observe from Fig.~\ref{fig4} that, in the case of flow-aligned pillars, the invasion occurs isotropically in the two-dimensional space, forming a rough front that eventually leads to $\mathcal{C}(t)  \sim \mathcal{C}_0$ and then the invasion dynamic stops, as predicted by Eq.~\eqref{eq:darcy2}. On the contrary, when the pillars are flow-opposing oriented, the invasion occurs preferentially along the transverse direction, the front is rather compact and  $\mathcal{C}(t) \ll \mathcal{C}_0$ so that the invasion events take place until the viscous fluid eventually reaches the outlet.

\section{The microstructural design of the porous capillary valve}

The design of the simple apparatus presented in Fig.~\ref{fig1} is thus motivated by the analysis presented in the previous sections. We chose a medium, with a longitudinal size $M_\parallel=4$, as thin as possible to obtain a fast invasion with flow-opposed pillars, but also thick enough to allow the front roughening along the other direction of injection. We placed the solid triangular pillars regularly in order to mimic the fabrication of a microfluidic system. The characteristic capillary threshold is again $p_c$ and we trigger a similar front roughening. We arbitrarily chose a percentage of defects $\zeta=0.06$. As a last expedient, we set the longitudinal characteristic length, i.e. the longitudinal distance between two defects belonging to the same pore column, as $1/(\zeta\ell_d)=4$, a value sufficiently high to prevent the formation of preferential channels along contiguous defects in the streamwise direction.

As we have discussed in the previous Section~\ref{sec:micro} the quantity $\mathcal{C}_0=\Delta p_0/p_{c}$ measures the maximum roughness achievable by the system. We also have observed that when the pillars are flow-aligned, the front roughness achieves a value $\mathcal{C}\sim 2$  at long times, while for the flow-opposing oriented pillars $\mathcal{C}\sim 1$.
Thus, following Eq.~\eqref{eq:darcy2}, for low values of the abscissa in Fig.~\ref{fig1}, i.e. for $\mathcal{C}_0\le 1$, the invasion process is  impeded after the first inertial stage, irrespective of the medium orientation. For intermediate values, $1<\mathcal{C}_0 \le 2$, the fluid is able to reach the outlet only for the medium orientation that keeps the front flat, which reduces the effective capillary resistance (flow-opposing pillars, $\mathcal{C} \sim 1$), while for larger values, i.e. $\mathcal{C}_0>2$, the fluid reaches the outlet irrespective of the medium orientation.

\section{Conclusions}

By means of Lattice Boltzmann two-phase flow simulations, we have investigated the possibility of achieving asymmetric flow conductance within porous media, through the application of anisotropic microstructural design. We have shown how the change in pore morphology, which yields anisotropy at the pore scale, together with the introduction of design defects, allows the creation of porous systems with a twofold functioning: if a viscous fluid is injected along a certain direction with respect to the medium orientation, it passes through the medium reaching the outlet, while it is blocked within the medium  when injected along the opposite direction. Such an asymmetric mechanism of transport can be a desired feature for the fabrication of porous capillary valves, devoted to the directional-dependent control of the flow.

The microstructural design of the porous medium that allows the asymmetric flow conductance consists of triangular-shaped pillars with randomly or regularly placed defects. Through the analysis of results of numerical simulations, we have assessed that such a specific configuration of the microstructure induces an invasion dynamics that depends on the medium orientation with respect of the direction of injection. When the triangular pillars are flow-aligned, the microscopic capillary thresholds determining the probability of invasion are isotropically distributed. As a consequence, the probabilities of invasion along the streamwise and transverse  directions are rather balanced. We have confirmed such an observation through the quantification of the distributions of streamwise and transverse pressure gradients, which, in the case of flow-aligned pillars, result similar along the two directions. Thus, the fluid invades isotropically the two-dimensional space, forming ramified fluid structures triggered by defetcs and a rough two-phase interface. We have also clarified that the higher the roughness of the two-phase interface, the higher the effective capillary resistance acting adversely to the flow, because it is larger the number of pore throats subjected to the microscopic capillary pressures related to the characteristic capillary threshold of the system $p_c$. The rough nature of the interface with this medium orientation, measured through the dimensionless parameter $\mathcal{C}\sim 2$, limits the invasion dynamics at long times.

On the contrary, when the triangular pillars are flow-opposing oriented, the probability of invasion along the transverse direction is higher than the one along the streamwise direction. We have indeed observed lower pressure gradients acting along the transverse direction, a measure of the presence of a large number of unstable fluid-fluid displacements at the pore scale and higher probabilities of invasion events. The invasion dynamics with such a medium orientation is thus following a serpentine pattern, mainly invading transversally all pores belonging to a row and, just in rare cases when the interface encounters a defect, invading along the streamwise direction. Such a mechanism of invasion dynamics keeps the invading front flat, minimise the two-phase interface roughness, measured through the dimensionless parameter $\mathcal{C}\sim 1$, and allows a more probable invasion of the viscous phase into the medium, until the fluid reaches the outlet.  In the last part of the paper, we have translated such a concept into a Darcy-like model that predicts the asymmetric permeability and flow conductance, on the basis of the structure and roughness of the invading viscous phase. The asymmetric functioning of the capillary valve is predicted for values of the ratio between maximum pressure drop and characteristic capillary threshold contained in the range $1 < \Delta p_0/p_c < 2$.

We anticipate similar but possibly more complex invasion patterns in three-dimensional
anisotropic materials. Having in mind that transverse invasion would then take place on a two-dimensional plane, we can expect an even slower invasion dynamics along such a direction~\cite{vaitukaitis2020water}. Future simulations should investigate the impact of three-dimensional invasion dynamics on the extent of the asymmetric functioning range currently observed. Furthermore, to elucidate the effect of random orientations of the pillars would be desirable in order to predict the behaviour of more realistic materials.

\begin{acknowledgments}
The research project to which this work belongs has received funding from the European Union's Horizon 2020 research and innovation programme under the Marie Sklodowska-Curie grant agreement No 790744. This work has been also supported by the Swedish Research Council for Environment, Agricultural Sciences and Spatial Planning (FORMAS), grant number 2019-01261.
The computations were enabled by resources provided by the Swedish National Infrastructure for Computing (SNIC) at C3SE and HPC2N partially funded by the Swedish Research Council through grant agreement no. 2018-05973. 
\end{acknowledgments}

\bibliography{apstemplate}

\end{document}